\documentclass[aps,prl,amsmath,amssymb,twocolumn,showpacs,superscriptaddress]{revtex4-1}

\usepackage{graphicx}
\usepackage{dcolumn}
\usepackage{bm}
\usepackage{color}
\usepackage{subfigure}
\usepackage{sidecap}
\usepackage[normalem]{ulem}

\newcommand {\apgt} {\ {\raise-.5ex\hbox{$\buildrel>\over\sim$}}\ }
\newcommand {\aplt} {\ {\raise-.5ex\hbox{$\buildrel<\over\sim$}}\ }

\draft 

\begin{document}


\title{Observation of Rayleigh--Taylor-instability evolution in a plasma with magnetic and viscous effects}

\author{Colin S. Adams}\thanks{Now at Virginia Polytechnic Institute and State University, Blacksburg, VA\, 24061.}
\affiliation{Los Alamos National Laboratory, Los Alamos, NM  87545}
\affiliation{University of New Mexico, Albuquerque, NM 87131}
\author{Auna L. Moser}\thanks{Now at General Atomics, San Diego, CA\, 92186.}
\author{Scott C. Hsu}
\email{Electronic mail: scotthsu@lanl.gov.}
\affiliation{Los Alamos National Laboratory, Los Alamos, NM 87545}

\date{\today}

\begin{abstract}

We present time-resolved observations of Rayleigh--Taylor-instability (RTI) evolution at the interface between an
unmagnetized plasma jet colliding with a stagnated, magnetized plasma.  The observed instability
growth time ($\sim 10$~$\mu$s) is consistent with the estimated linear RTI growth rate calculated using
experimentally inferred values of density ($\sim 10^{14}$~cm$^{-3}$) and deceleration ($\sim 10^9$~m/s$^2$).
The observed mode wavelength ($\gtrsim 1$~cm) nearly doubles within a linear growth time.
Theoretical estimates of magnetic and viscous stabilization and idealized magnetohydrodynamic simulations
including a physical viscosity model both suggest that the observed instability evolution
is subject to magnetic and/or viscous effects.



\end{abstract}



\maketitle 

Rayleigh--Taylor instabilities (RTI) \cite{rayleigh_1883, taylor_1950, chandrasekhar_1961} have been well-studied in neutral fluids for more than a century.  RTI also plays an important role in astrophysical and laboratory plasmas, e.g., possibly in the formation of the pillars in the Eagle nebulae \cite{kane_AAS} and in all imploding fusion systems, e.g.,
\cite{atzeni04textbook,kirkpatrick95ft,ryutov00rmp}.
In a plasma, both magnetic field and a strongly temperature-dependent viscosity can dramatically affect the evolution of RTI \cite{chandrasekhar_1961, harris62pof} and its consequences on plasma transport \cite{srinivasan13pop,srinivasan14epl}.
While study of RTI in the context of inertial confinement fusion (ICF) spans many decades, e.g., \cite{atzeni04textbook} and references therein,
only more recently have studies seriously focused on the effects of magnetic field or viscosity on RTI evolution in
plasmas, e.g., \cite{sze07pop,sinars_2010,osin11ieee, mcbride12prl, srinivasan13pop,mcbride13pop,gomez14prl,srinivasan14epl}.
Detailed experimental data on the latter
are especially needed for validating simulations of
magnetized ICF \cite{hohenberger12pop,perkins13pop,srinivasan14epl} and
magneto-inertial fusion (MIF) \cite{slutz10pop,degnan13nf}.

In this work, we present time-resolved observations of unseeded RTI growth and evolution at a decelerating, mostly planar
plasma interface,
focusing on the role played by magnetic field and viscosity in RTI evolution and stabilization.
Our plasmas have spatial scales of tens of cm, temporal durations of tens of $\mu$s, and density $\sim 10^{14}$~cm$^{-3}$.  These factors
allow us to experimentally measure or infer interfacial plasma parameters, including velocity $v$,
local magnetic field $B$, electron density $n_e$ and temperature $T_e$, mean-ionization state $\bar{Z}$,
ion viscosity $\nu$ (assuming $T_i=T_e$), and to obtain time-resolved measurements of RTI growth in a single shot.  The key results are:  (1)~identification of instability growth consistent with theoretical estimates
of linear RTI, (2)~observation of the evolution toward longer instability mode wavelengths over a linear
growth time,
and (3)~analyses suggesting that the observed instability evolution is subject to 
magnetic and/or viscous effects.



\begin{figure}[!tb]
\begin{center}
\includegraphics[width=8.6cm]{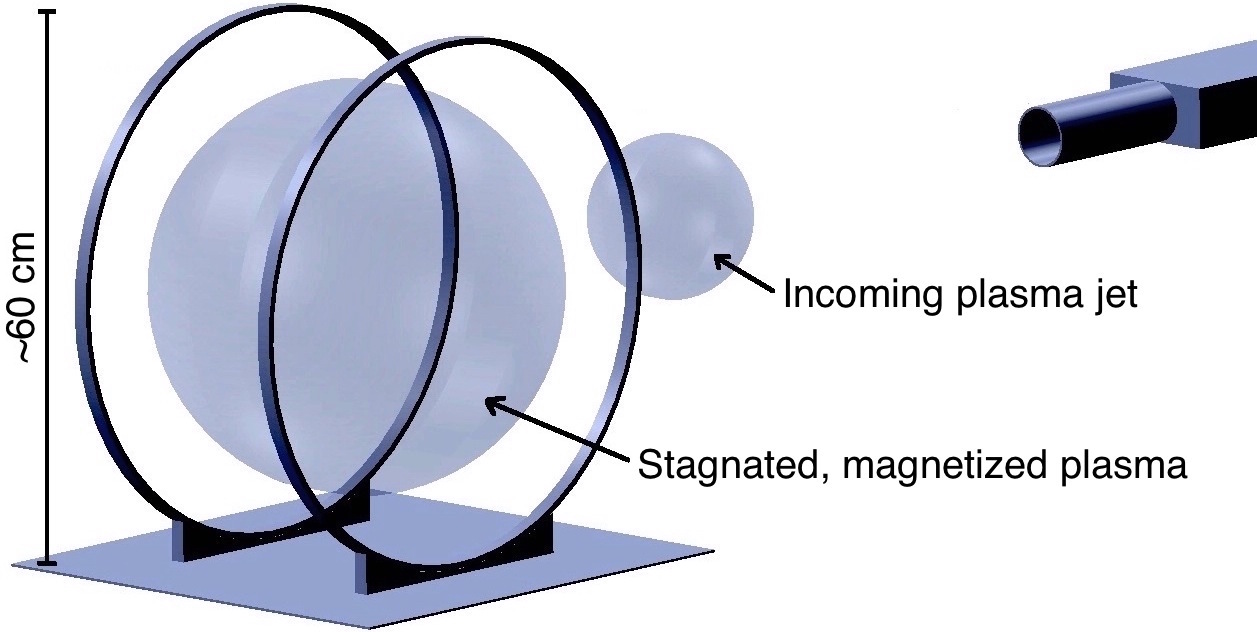}
\caption{\label{f:exp} (Color online) Experimental setup: two plasma jets launched by oppositely
positioned railguns (only one shown at right) collide head-on near the
center of an in-chamber HH coil, giving rise to a stagnated plasma.  We study the interaction of
a second jet (formed due to the ringing railgun current) colliding with the stagnated plasma.}
\end{center}
\end{figure}

Experiments presented here were conducted on the Plasma Liner Experiment \cite{hsu12ieee, hsu12pop, hsu14jpp}.  Two plasma-armature railguns fire plasma jets (composed of an
argon/impurity mixture) that merge head-on after each propagating approximately 1.1 meters to the center
of a 9-ft.-diameter spherical vacuum chamber.  Previously, individual jets \cite{hsu12pop},
two obliquely merging jets \cite{merritt13prl, merritt14pop}, and head-on merging jets \cite{auna2014dpp} have been well characterized.  In this work, we have added Helmholtz (HH) coils at the center of the vacuum chamber to produce a magnetic field perpendicular to the jet-propagation direction \cite{hsu14jpp}.  
The HH current has a rise-time of $\approx 1.3$~ms and is thus essentially steady-state on the time scale of the
interaction ($\sim 10$~$\mu$s). Due to underdamped ringing of the railgun electrical current,
plasma jets are released from each gun in a series with 30-$\mu$s intervals.  After the leading
jets from each gun collide and form a stagnated plasma within the applied magnetic field,
the second jet arrives in the interaction region (Fig.~\ref{f:exp}).  
Here we focus on the interaction of the second jet (from one of the guns) with the stagnated plasma.
A fast-framing camera (Invisible Vision UHSi 12/24), magnetic probe array, survey spectrometer, multi-chord interferometer \cite{merritt12rsia, merritt12rsib}, and photodiode array are employed to study the interaction region.  
The $n_e$, $T_e$, and $\bar{Z}$ are determined
via an iterative data-analysis process \cite{hsu12pop} that utilizes interferometry and spectroscopy data and non-local-thermodynamic-equilibrium PrismSPECT \cite{prism} spectral calculations.  Plasma velocity is estimated via
time-of-flight of features in both the photodiode-array and interferometer signals.  
Magnetic field values at a single
spatial position are determined by integrating signals from magnetic pickup coils inserted into the chamber. More diagnostic details are given elsewhere \cite{hsu12pop, hsu14jpp}. 

\begin{figure*}[!htb]
\includegraphics[width=17.8cm]{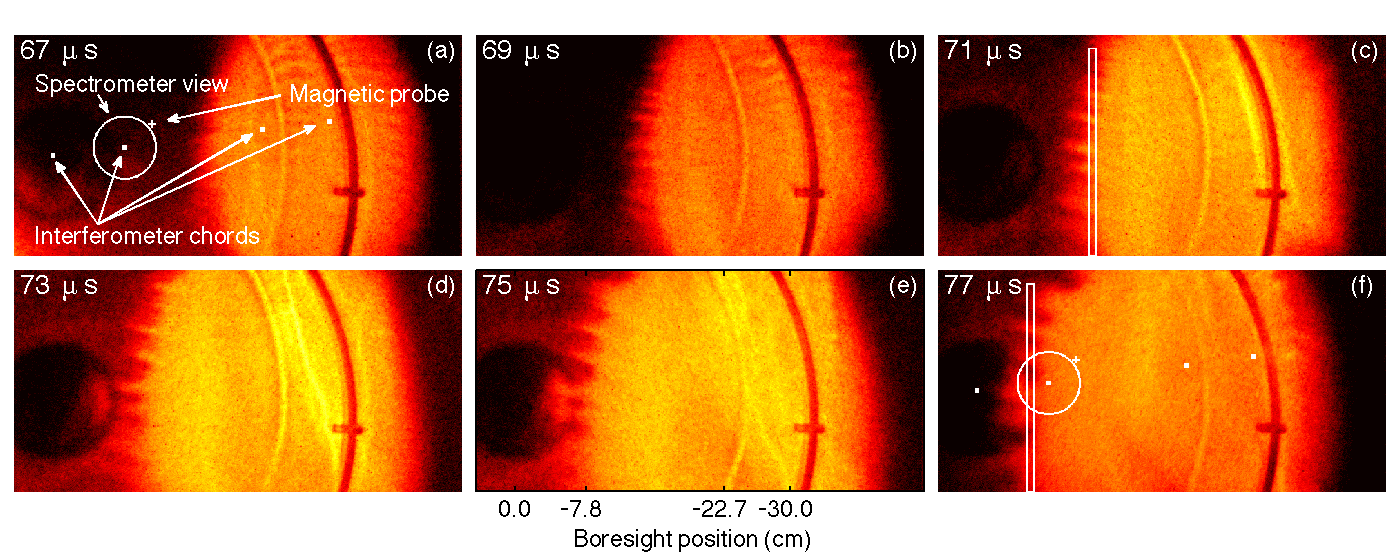}
\caption{\label{f:seq} (Color online) Fast-camera images (from a single shot, 3159) capturing instability growth as the second jet (brighter region, moving from right to left) interacts with the stagnated plasma (darker region on the left).  The location of diagnostic measurements are shown in (a) and (f).  The arc-shaped structures visible in the right side of each image are the HH coils. The view is nearly parallel to the applied magnetic field at the center of the HH coils.  Interferometer chord positions 
shown are at 0, -7.5, -22.5, and -30~cm along the boresight axis, which is aligned with the railgun bores; the magnetic
probe is at $\approx -10$~cm.}
\vskip-2ex

\end{figure*}

Figure~\ref{f:seq} shows fast-camera images (from a single shot)
of the interaction of the second jet with the stagnated plasma.  
The image resolution at the center of the Helmholtz coils is
$\approx 750$~$\mu \mathrm{m}/$pixel, substantially smaller than any observed features.  Line-of-sight (if 
the instability wave vector rotates in time) or motion blur (during the 750-ns 
exposure time) could contribute to loss of small-scale detail.
The stagnated plasma is dark and centered in the vicinity of the spectrometer view, which is near the center of the HH coils
($\approx 30$-$\mathrm{cm}$ radius).  As the second jet impacts the stagnated plasma, the jet slows down.  RTI-like fingers are easily seen in the images.
Tracking the interface location in the images of Fig.~\ref{f:seq} indicates that the interface slows from $\approx16$ to 
$\approx 7$~$\mathrm{km/s}$ between $t=69$--77~$\mu\mathrm{s}$, corresponding to a deceleration of $\approx 1 \times 10^{9}$~$\mathrm{m/s^2}$.

\begin{figure}[h]
\begin{center}
\includegraphics[width=7.6cm]{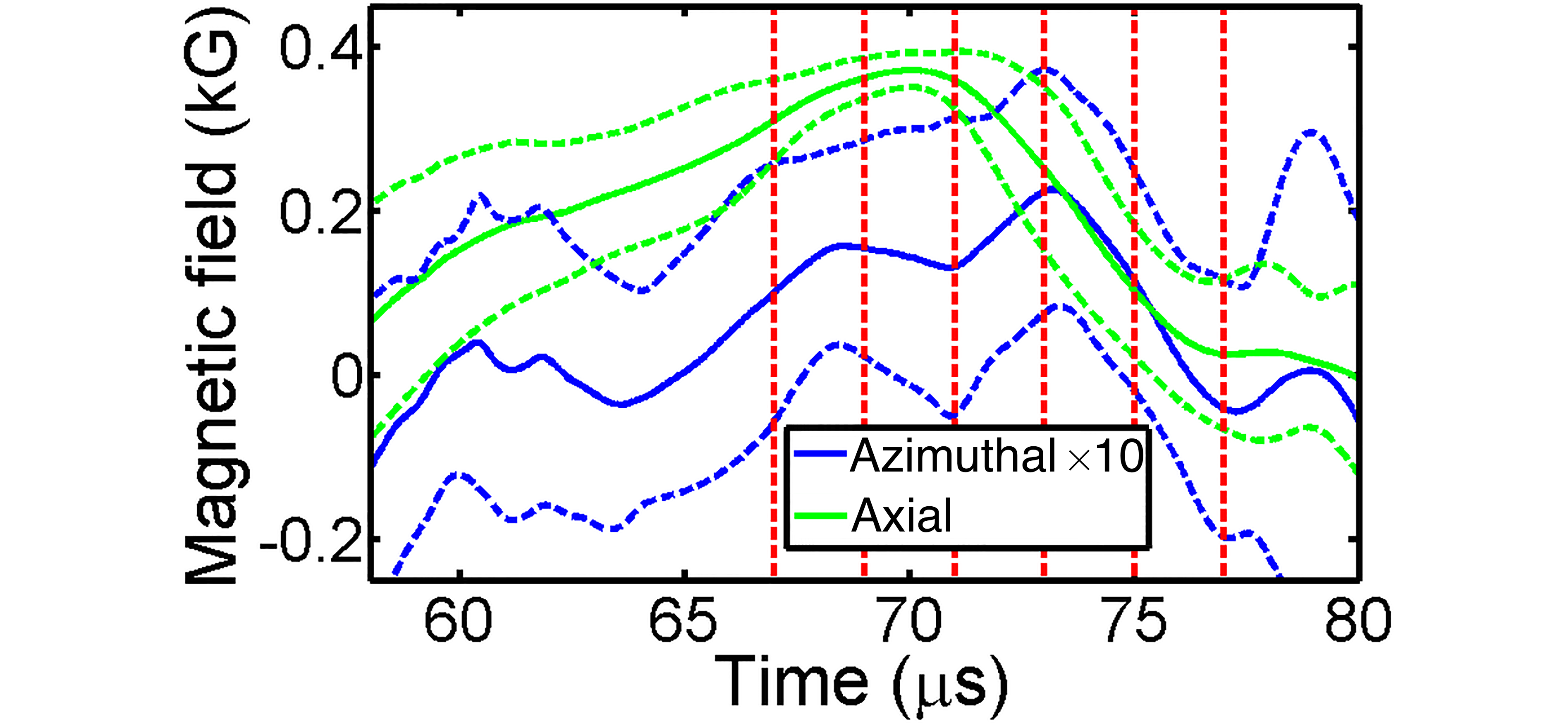}
\caption{\label{f:mag} (Color online) Axial and azimuthal$\times 10$ magnetic field vs.\ time 
(averaged over subset of shots 3209--3232; dashed lines indicate standard deviation)
as measured by the $\dot{B}$ probe 
added to the applied field.  The coordinate system is with respect to the HH-coil axis, i.e., ``axial" is approximately into 
the page in Fig.~\ref{f:seq}.  The radial component (not shown) shows much less activity than the axial component 
during the time of interest.  The vertical dashed lines indicate the beginning of each exposure in Fig.~\ref{f:seq}.}
\end{center}
\end{figure}

We use an in-chamber magnetic probe (position shown in Fig.~\ref{f:seq}) to measure the local magnetic field 
$\vec{B}$ during the interaction of the second jet with the stagnated plasma (Fig.~\ref{f:mag}), showing that
the field strength in the HH coil's axial direction (perpendicular to the jet-propagation direction) grows to $B_{\rm axial} >300$~G and then falls dramatically to $\approx 0$~G 
and in the coil's azimuthal direction (nearly aligned with the mode wavevector $\vec{k}$) to be $B_{\rm azim}\approx 15$~G
during the time range shown in Fig.~\ref{f:seq}.
For the shot shown in Fig.~\ref{f:seq}, the HH coil had a 1.1-kA peak current corresponding to an $\approx 290$-G
applied field along the HH axis.
If the HH coils are not energized, the field is below the
detection limit of our probe ($\approx 10$~G).  Thus, the second jet is considered to be unmagnetized as it arrives in the vicinity of the HH coils.  

\begin{figure}[h]
\begin{center}
\includegraphics[width=8.0cm]{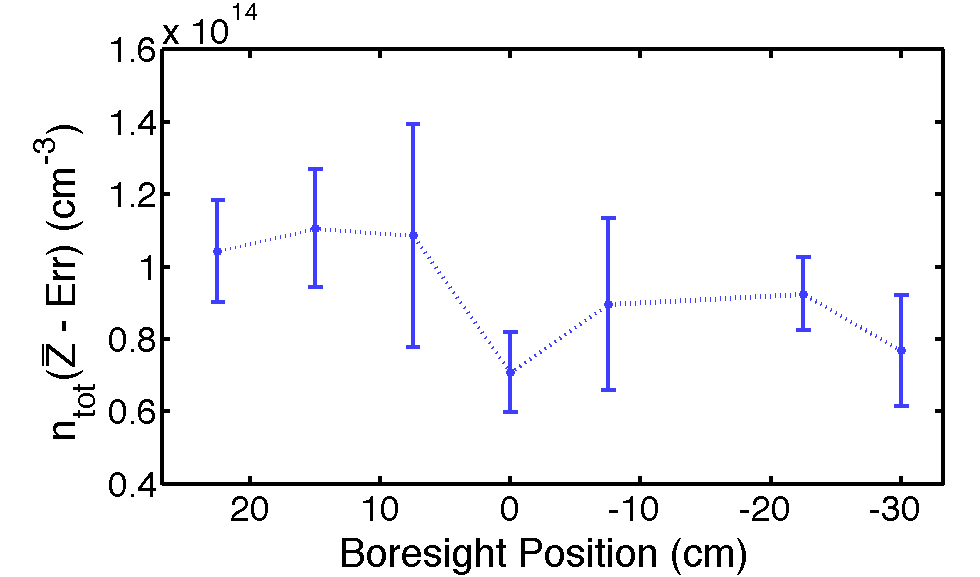}
\caption{\label{f:interf} (Color online) Chord-averaged density $n_{tot}\approx n_e/\bar{Z}$ at $t=77$~$\mu 
\mathrm{s}$
(averaged over subset of shots 3153--3190), using a chord length of 30~cm estimated from camera images, along seven interferometer chords.   Error bars indicate standard deviation over multiple shots. }
\end{center}
\end{figure}

Seven interferometer chords are used to measure the spatial and temporal evolution of phase shift $\Delta \Phi$ from free and bound electrons in the plasma \cite{merritt12rsia, merritt12rsib}, with chord-integrated density
$\int n_{\mathrm{tot}} dl = \Delta \Phi / \left[ C_e \left( \bar{Z} - Err \right) \right]$, where $n_{\mathrm{tot}}$ is
the total ion-plus-neutral density, $C_e = (\lambda_\mathrm{probe} e^2)/(4 \pi \epsilon_0 m_e c^2) = {1.58 \times 10^{-21}}$~$\mathrm{rad \cdot m^2}$ is the phase sensitivity to electrons, $\lambda_{\mathrm{probe}} = 561.3$~nm, and $Err \lesssim 0.08$ is a bound on the contribution from bound electrons \cite{merritt14pop}, which for cases of interest is small compared to $\bar{Z}$.  Figure~\ref{f:interf} shows the spatial profile of $n_{\mathrm{tot}} \left( \bar{Z}- Err \right)$ (with an assumed chord length of 30~cm), from which we infer that $n_{\mathrm{tot}} \sim 10^{14}$~$\mathrm{cm^{-3}}$.

\begin{figure}[h]
\begin{center}
\includegraphics[width=8.6cm]{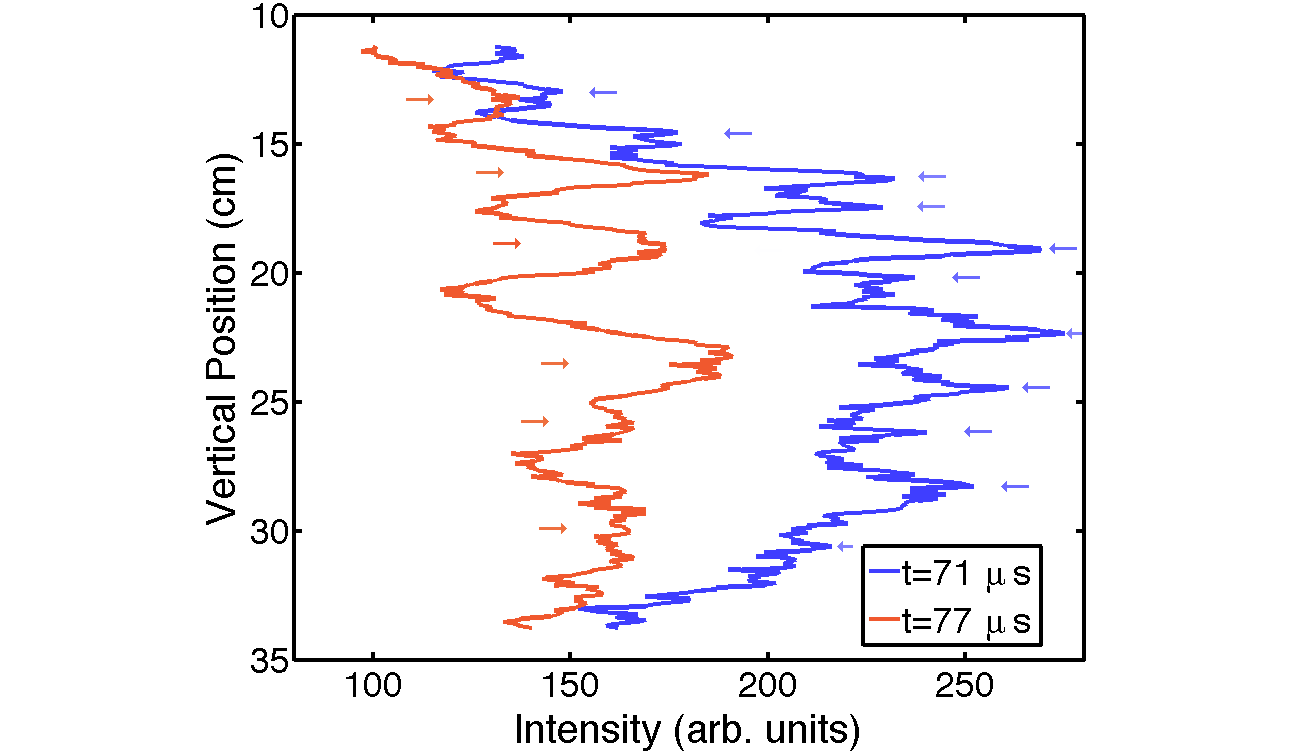}
\caption{\label{f:spect} (Color online) Summed emission lineouts from Figs.~\ref{f:seq}(c) and (f).  Pixel intensity of 10 adjacent (300-pixel high) columns are summed together across the regions of each frame containing the observed fingers.  The locations of the bright fingers are highlighted with arrows.  The instability wavelength decreases by a factor of $\approx 5/3$ over 6~$\mu$s.}
\end{center}
\end{figure}

The images in Fig.~\ref{f:seq} show an increase in instability wavelength versus time.  To quantify this, 
we plot the sum of vertical lineouts (Fig.~\ref{f:spect}) from the regions indicated in Figs.~\ref{f:seq}(c) and (f).  The lineouts  show that 10 fingers appear at $t=71$~$\mu$s, while only 6 fingers appear at $t=77$~$\mu$s in a region of the same height.  The displayed length scales are uncorrected for parallax, but this effect is negligible because the camera is situated approximately two meters from the jet and the camera line-of-sight is nearly perpendicular to jet propagation.  The $5/3\times$ increase in wavelength over $6$~$\mu$s suggests that small wavelengths are
being stabilized.

\begin{figure}[!tb]
\begin{center}
\includegraphics[width=6.0cm]{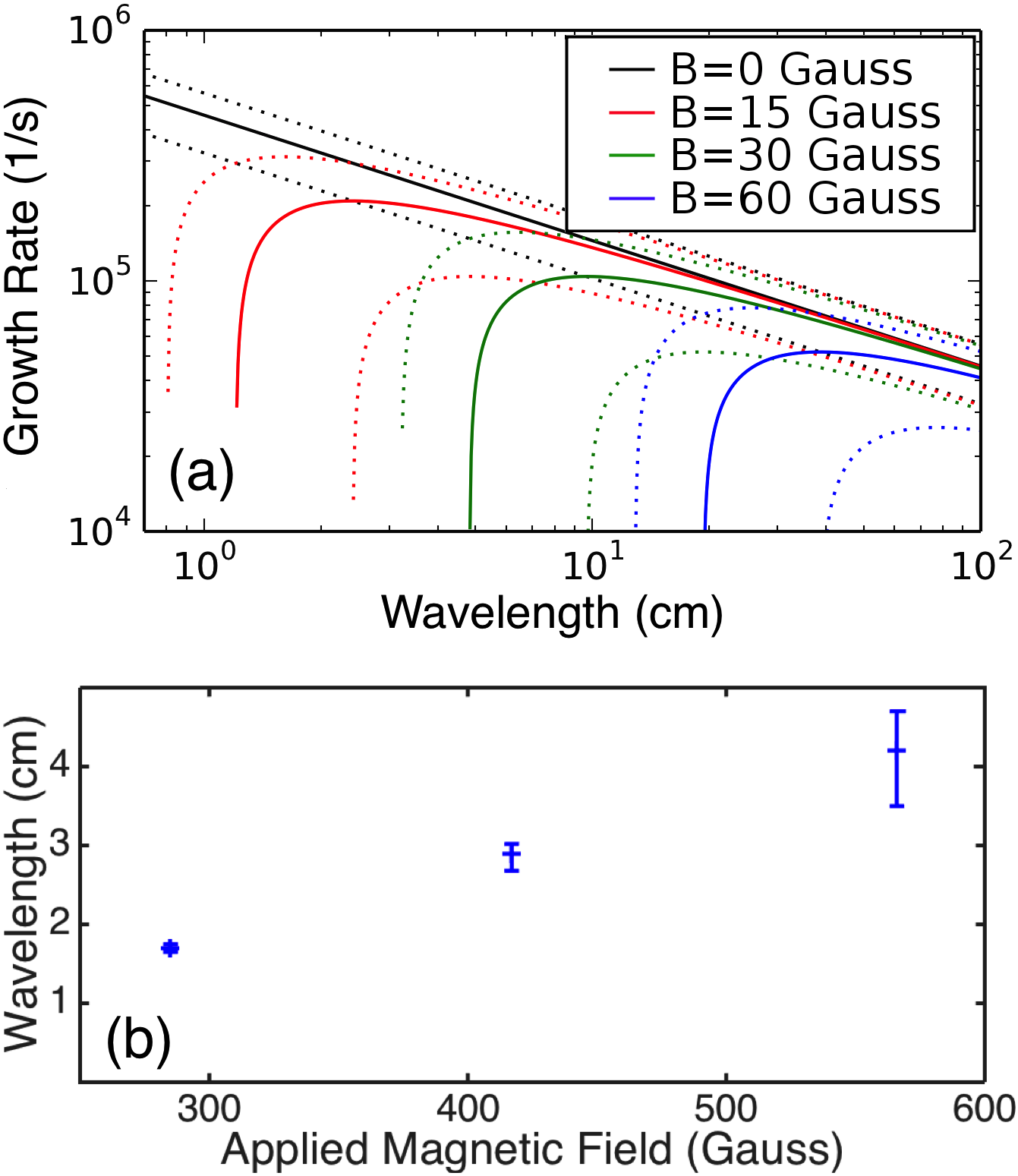}
\caption{\label{f:grow} (Color online) (a)~Calculated linear growth rate vs.\ magnetic-RTI wavelength for different $B_\|$ relevant
to our experiments.  Dashed lines correspond to $A \pm 50$\%, showing the relatively small effect of
uncertainty in $A$.  (b)~Wavelength of observed instabilities as a function of applied $B_{\rm axial}$ (shots 
3159, 3186, 3182).  Error bars indicate uncertainty in interpreting emission lineouts.}
\end{center}
\end{figure}

First, we consider magnetic stabilization.  Linear magnetic-RTI theory \cite{chandrasekhar_1961, harris62pof}
predicts a growth rate $\gamma=\sqrt{gkA - (\vec{k} \cdot \vec{B})^2/[2 \pi (\rho_2 + \rho_1)]}$
(cgs units),
where $g$ is the acceleration, $\rho_1$ and $\rho_2$ are mass densities on either side of the interface,
and $A=(\rho_2-\rho_1)/(\rho_2+\rho_1)$ is the Atwood number.
Figure~\ref{f:grow}(a) shows $\gamma$ versus RTI wavelength for $B_\| = \vec{B}\cdot \vec{k}/k$ values relevant to
our experiment (using $\rho_2 = 2  \rho_1 \approx 5.6 \times 10^{-9}$~$\mathrm{g/cm^3}$).
The experimentally observed wavelengths are $\approx 2$~cm with growth time $\lesssim 10$~$\mu$s, which would be consistent
with the measured $B_{\rm azim} \approx 15$-G (near-vertical direction in Fig.~\ref{f:seq}).
However, we measured $\vec{B}$ at only one spatial position, and the overall magnetic structure is unknown.  If the
field is spatially uniform on the scale of the mode structure, then the mode is likely along the
$\vec{k}\cdot \vec{B}=0$ direction ($\approx \arctan[(B_{\rm azim}=15~\mathrm{G})/(B_{\rm axial}=290~\mathrm{G})] \approx 3^\circ$ from vertical) with little magnetic stabilization but possibly viscous stabilization.
If, on the other hand, the field
is varying (and/or sheared) on the scale of the mode structure (as suggested by the relatively large
shot-to-shot
standard deviation in $B_{\rm azim}$ in Fig.~\ref{f:mag}), then the observed mode could be the
net result of stabilization by a nonuniform, dynamic field structure.
Due to the plasma jet's kinetic energy density being $\approx 30$ times greater than its magnetic pressure,
it is plausible that the observed $B_{\rm azim}$ results from advection leading to reorientation of the applied 
$B_{\rm axial}$.
Additional argon experiments with applied $B_{\rm axial} \approx 420$--570~G
show approximately linear scaling of the observed instability wavelength with applied $B_{\rm axial}$
[Fig.~\ref{f:grow}(b)], further suggesting that the observed mode evolution is subject to magnetic effects.

Next, we consider plasma viscosity, which stabilizes wavelengths shorter than a maximum wavelength
$\lambda_\mathrm{max} = 4 \pi [\nu^2 /(g A)]^{1/3}$,
where $\nu$ is the kinematic viscosity \cite{plesset_1974}.
In our case, the viscosity is dominated by the ion viscosity due to their much larger gyro-radii than electrons,
in which case viscosity is proportional to $T_i^{5/2} \bar{Z}^{-4}$, and thus $\lambda_\mathrm{max} \sim T_i^{5/3} \bar{Z}^{-8/3}$.  Because the jets are initially collisional with ion-electron energy-equilibration time of $\approx 30$~$\mu\mathrm{s}$, it is reasonable to assume that $T_i \approx T_e$ in the second jet.  
Comparing PrismSPECT spectral calculations with experimental spectra collected over a series of shots
covering times both before and after the interface passes the
location of the spectrometer enables bounding of $T_e$ in both the stagnated plasma and second jet.  Prior to the arrival of the second jet, the appearance of line emission near 497.2~nm and the lack of line emission at 520.8~nm indicate a peak $T_e \approx 2.3$--2.4~eV in the stagnated plasma.  After the interface passes the spectrometer view (corresponding to boresight-position $-7.5$~cm), the appearance of line emission near 490.6~nm and the lack of line emission at 453.1~nm indicate a peak $T_e \approx 2.7$--2.8~eV in the second jet.  
Examples of spectra and comparisons with PrismSPECT 
spectral calculations have been presented elsewhere for similar experiments \cite{hsu12pop,merritt13prl} and thus are not shown here.
For the range $2.3 < T_e < 2.8$~eV and corresponding $1.2 < \bar{Z} < 1.6$ (from PrismSPECT calculations),
dynamic viscosities in the range $5.2\times 10^{-5} < \mu= \rho \nu~< 1.1\times 10^{-4}$ $\mathrm{g/(cm \cdot s)}$ are possible.
These viscosities give $\lambda_{\rm max} \approx$ 1.7--2.9~cm (using $g= 10^9$~$\mathrm{m/s^2}$, $\rho_2 = 2  \rho_1 = 5.6 \times 10^{-9}$~$\mathrm{g/cm^3}$), which is consistent with the observed wavelengths.  

To further explore magnetic and viscous effects in our experimental regime, two-dimensional simulations of RTI growth were computed using WARPX \cite{srinivasan2011ccp, shumlak_2011, srinivasan14epl}, in which an ideal magnetohydrodynamic (MHD) model is solved with a discontinuous Galerkin method.  An interface between two regions (with $\rho_2 = 2 \rho_1 = 6 \times 10^{-9}$~$\mathrm{g/cm^3}$) is simultaneously perturbed with three different wavelength seeds of 1, 4, and 20~cm.  Computational resolution is 1~mm and numerical viscosity is negligible for the wavelengths studied.  Subjecting the perturbed interface to an acceleration of $10^{9}$~$\mathrm{m/s^2}$ causes RTI growth.  An array of different initially uniform magnetic field strengths and viscosities were compared to isolate and assess the effects of magnetic field and viscosity.

\begin{figure*}[!tb]
\includegraphics[width=17.8cm]{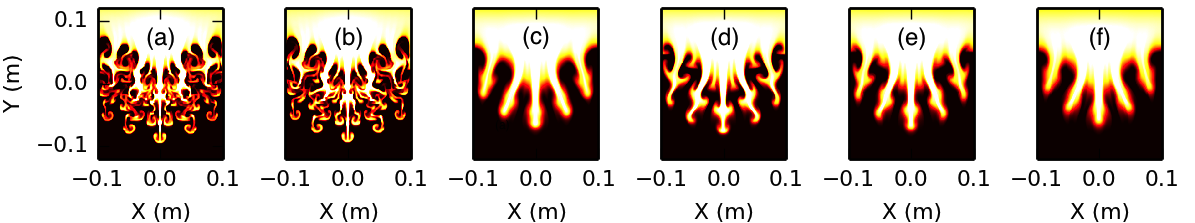}
\caption{\label{f:sim} (Color online) Ideal 2D MHD simulation of RTI evolution at a plasma interface.   Case with no magnetic field (a) and case with magnetic field $B_x=2$~G (b) do not show stabilization of any modes after 30~$\mathrm{\mu s}$ of growth, while cases with magnetic field $B_x=15$~G (c), ion viscosity of $5.2\times10^{-5}$~$\mathrm{cm^{-1}s^{-1}g}$ (d), or ion viscosity of $1.1\times10^{-4}$~$\mathrm{cm^{-1}s^{-1}g}$ (e) show stabilization of short-wavelength RT modes.  A case with viscosity of $1.1\times10^{-4}$~$\mathrm{cm^{-1}s^{-1}g}$ and magnetic field $B_x=15$~G (f) shows similar growth characteristics as solely magnetic stabilization. }  
\vskip-2ex
\end{figure*}

Figure~\ref{f:sim} shows the simulation results after 30~$\mathrm{\mu s}$ of growth for six cases, including two magnetic field magnitudes (aligned along $\vec{k}$) and two values of viscosity.  Figure~\ref{f:sim}(a) shows small-scale mode growth in the absence of both $B$ and $\mu$.  While these small-scale structures would be expected to cool more rapidly than longer mode wavelengths \cite{srinivasan14epl}, perhaps affecting their experimental observability, our computations did not utilize an emission model. 
Figures~\ref{f:sim}(b) and (c) with zero physical viscosity show that a horizontal magnetic field of 2~G is incapable of stabilizing even 1-cm modes, while a field of 15~G is capable of stabilizing 1-cm but not 4-cm modes.  Figures~\ref{f:sim}(d) and (e) have no magnetic field but have viscosities corresponding to $T_e= 2.8$~eV, $\bar{Z} = 1.6$ and $T_e=2.3$~eV, $\bar{Z} = 1.2$, respectively.  While both cases are capable of stabilizing 1-cm modes, the 2.8-eV case does so poorly, while the 2.3-eV case is qualitatively similar to the 15-G case.  Finally,
Fig.~\ref{f:sim}(f) shows a simulation with both high viscosity and a 15-G magnetic field, resulting in evolution similar to the 15-G field case.  These simulations further suggest that the observed instability evolution is
subject to magnetic and/or viscous effects.

In  summary, we presented time-resolved experimental observations of RTI evolution at a decelerating
plasma interface in the presence of an applied magnetic field.
The observed instability wavelength ($\approx 2$~cm) and growth time ($\sim 10$~$\mu$s) are consistent
with linear RTI theory using experimentally inferred values of density and deceleration.
In addition, we observed a near doubling of the mode wavelength within
one linear growth time and an approximately linear scaling of the observed mode wavelength with applied field.
Theoretical estimates and idealized MHD simulations both suggest that the observed RTI evolution is subject to magnetic and/or viscous effects in our parameter regime.
The impact on instability growth of non-uniform magnetic field, thermal conduction,
and resistive effects at the interface, as well as comparisons between experimental
and synthetic diagnostic data, should be investigated in future work. These experimental data can aid the validation
of models used to simulate, e.g., the mitigation of mix in ICF by magnetic
and viscous stabilization \cite{srinivasan14epl},
and are relevant to the study of MIF-relevant deceleration-phase magnetic RTI\@.

\begin{acknowledgments}
We acknowledge J. Dunn for experimental support, G. Wurden for loaning and assisting with the multiple-frame CCD camera, and B. Srinivasan for assistance with simulations.  This work was supported by the Laboratory Directed Research and Development (LDRD) Program at Los Alamos National Laboratory under DOE contract no.\ DE-AC52-06NA25396.
\end{acknowledgments}


%
%

%


%

\end{document}